\documentclass[aps,prl,twocolumn,superscriptaddress,showpacs]{revtex4}

\usepackage{graphicx,amsmath}

\newcommand{\rv}{{\bf r}}
\newcommand {\dr}{{\mathrm d}{\bf r}}

\begin{document}

\title{Dynamics in inhomogeneous liquids and glasses via the test particle limit}

\author{Andrew J.\ Archer} \affiliation{H.H.\ Wills Physics
Laboratory, University of Bristol, Tyndall Avenue, Bristol BS8 1TL,
UK}

\author{Paul Hopkins}
\affiliation{H.H.\ Wills Physics Laboratory, University of Bristol,
Tyndall Avenue, Bristol BS8 1TL, UK}

\author{Matthias Schmidt}
\affiliation{H.H.\ Wills Physics Laboratory, University of Bristol,
Tyndall Avenue, Bristol BS8 1TL, UK}
\affiliation{Institut f\"ur Theoretische Physik II,
Heinrich-Heine-Universit\"at D\"usseldorf, Universit\"atsstra\ss e 1,
D-40225 D\"usseldorf, Germany}

\date{26 September 2006}


\begin{abstract}
 We show that one may view the self and the distinct part of the van
 Hove dynamic correlation function of a simple fluid as the one-body
 density distributions of a binary mixture that evolve in time
 according to dynamical density functional theory. For a test case of
 soft core Brownian particles the theory yields results for the van
 Hove function that agree quantitatively with those of our Brownian
 dynamics computer simulations. At sufficiently high densities the
 free energy landscape underlying the dynamics exhibits a barrier as a
 function of the mean particle displacement, shedding new light on the
 nature of glass formation.  For hard spheres confined between
 parallel planar walls the barrier height oscillates in-phase with the
 local density, implying that the mobility is maximal between layers,
 which should be experimentally observable in confined colloidal
 dispersions.
\end{abstract}

\maketitle

The van Hove function $G(r,t)$ for the probability of finding a
particle at time $t$ at a distance $r$ from the origin, given that
there was a particle at the origin at time $t=0$, characterizes
dynamical phenomena in Condensed Matter on a nanoscopic
scale~\cite{vanhove54,Hansen06}. Recent measurements of $G(r,t)$ using
confocal microscopy were aimed at investigating e.g.\ dynamical
heterogeneities in colloidal hard sphere suspensions \cite{kegel00},
or the devitrification of colloidal glasses upon adding nonadsorbing
polymers \cite{simeonova06}.  The Fourier transform of $G(r,t)$, the
intermediate scattering function, can be measured with inelastic
scattering techniques~\cite{Hansen06}.

There is no consensus on a theoretical approach for investigating the
microscopic dynamics in dense {\em inhomogeneous} liquids, relevant to
studying e.g.\ the glass transition for liquids adsorbed in nanoporous
materials~\cite{alcoutlabi05}. This situation contrasts with the
static case, where modern density functional theory (DFT) provides a
powerful means for investigating the significant effects on phase
behaviour and structural correlations that spatial confinement and
external fields may induce \cite{Hansen06,Evans92}.  DFT operates on
the level of the one-body density distribution $\rho(\rv)$, where
$\rv$ is the space coordinate. For a range of situations the time
evolution of the density profile, $\rho(\rv,t)$, as induced by a
time-dependent external potential, $V_{\rm ext}(\rv,t)$, has been
shown to be well described by dynamical density functional theory
(DDFT) \cite{MarconiTarazona,ChanFinkenPRL2005}, see e.g.\ the
successful applications in Refs.\ \cite{MarconiTarazona,litgaussians,
Matthiasetal}. However, currently no similar theoretical framework
exists for calculating dynamical two-body correlation functions, as
required to address problems such as those mentioned above.

In the static case a close relationship between $\rho(\rv)$ and the
pair distribution function, $g(r)$, describing the probability of
finding a pair of particles separated by a distance~$r$, is
established by Percus' famous test particle limit \cite{percus62}:
$\rho g(r)$ is the {\em one-body} density distribution of a fluid
exposed to the influence of an external potential $V_\text{\rm
ext}(r)$, that represents a `test particle' fixed at the origin, given
by $V_\text{\rm ext}(r)=V(r)$, where $V(r)$ is the interparticle pair
potential, and $\rho$ is the bulk density.

In this Letter we generalize the test particle limit to dynamical
correlation functions allowing us to use DDFT to calculate the van
Hove function in bulk and in inhomogeneous systems. We test this
approach by comparing results to those from our Brownian dynamics
computer simulations for a simple Gaussian core model (GCM) for a
macromolecular solution in bulk, and indeed find very good
agreement. For hard spheres we estimate the glass transition to be at
bulk densities $\rho \sigma^3\simeq 0.9$, where $\sigma$ is the hard
sphere diameter. Upon confining the hard sphere system between
parallel hard walls, we find the local mobility to be maximal where
the local density is minimal. Our approach allows for inspection of a
well-defined underlying free energy landscape, which requires solving
a corresponding equilibrium DFT rather than the full DDFT, easing
numerical burdens.

Consider a fluid of $N$ particles that interact via a pair potential
$V(r)$. The system may be viewed as a binary mixture of species $s$
and $d$, in which two of the pair potentials $V_{ij}(r)$ between
particles of species $i,j=s,d$ are equal, $V_{dd}(r)=V_{sd}(r)=V(r)$,
and $V_{ss}(r)=0$. Furthermore species $s$ consists of only a single
(test) particle, located at position $\rv_0$ at $t=0$, while species
$d$ consists of the remaining $N-1$ particles of the system. Our aim
is to relate the one-body density distributions of this mixture,
$\rho_s(\rv,t)$ and $\rho_d(\rv,t)$, to the self and distinct part of
the van Hove function of the pure fluid, $G_s(\rv_0,\rv,t)$, and
$G_d(\rv_0,\rv,t)$, respectively. [Recall that $G(r,t)=G_s(r,t)+
G_d(r,t)$.]  We must choose suitable initial conditions for
$\rho_s(\rv,t=0)$ and $\rho_d(\rv,t=0)$. In the case of a bulk system
let the one-body density distribution of species $s$ at time $t=0$ be
$\rho_s(\rv,t=0)=\delta(\rv)$, where $\delta(\cdot)$ is the Dirac
delta function and we have chosen the coordinate system such that
$\rv_0=0$. Species $d$ is initially at equilibrium, i.e.\
$\rho_d(\rv,t=0)=\rho g(r)$. For subsequent times, $t>0$, we make the
identification $G_s(r,t)=\rho_s(r,t)$ and $G_d(r,t)=\rho_d(r,t)$.

There are two ways to proceed. One is to carry out computer
simulations on the level of particle coordinates (detailed below). The
alternative is to use a theory that operates on the one-body level,
and DDFT is an obvious choice. Generally there are only formal results
for the exact equations for the time evolution and hence one must
resort to approximations
\cite{MarconiTarazona,ChanFinkenPRL2005,Archer7and8,Archer17}, such as
the theory of Marconi and Tarazona \cite{MarconiTarazona} for Brownian
dynamics:
\begin{align}
  \frac{\partial \rho_i}{\partial t} &= \Gamma \nabla \cdot
  \left(\rho_i \nabla \frac{\delta F[\rho_s,\rho_d]}{\delta
    \rho_i}\right), \quad i=s,d,
  \label{eq:MT_DDFT}
\end{align}
where $\Gamma$ is a friction coefficient characterising the drag of
the solvent on the particles, and $F[\rho_s,\rho_d]$ is taken to be
the {\em equilibrium} Helmholtz free energy functional:
\begin{align}
  F[\rho_s,\rho_d] = &
  k_BT\sum_{i=s,d} \int \dr\rho_i(\rv)
      [\ln\left(\rho_i(\rv)\Lambda^3\right)-1]  \label{eq:F}\\ &
      \;+F_{\rm ex}[\rho_s,\rho_d]
      +\int \dr V_{\rm ext}(\rv) [\rho_s(\rv)+\rho_d(\rv)], \notag
\end{align}
where the first term on the r.h.s.\ is the Helmholtz free energy
functional of the (binary) ideal gas, $\Lambda$ is the (irrelevant)
thermal wavelength, $T$ is the temperature, $k_B$ is the Boltzmann
constant, and $F_{\rm ex}[\rho_s,\rho_d]$ is the excess contribution
to the Helmholtz free energy due to interactions between the particles
\cite{canonical}.

\begin{figure}
\includegraphics[width=0.45\columnwidth]{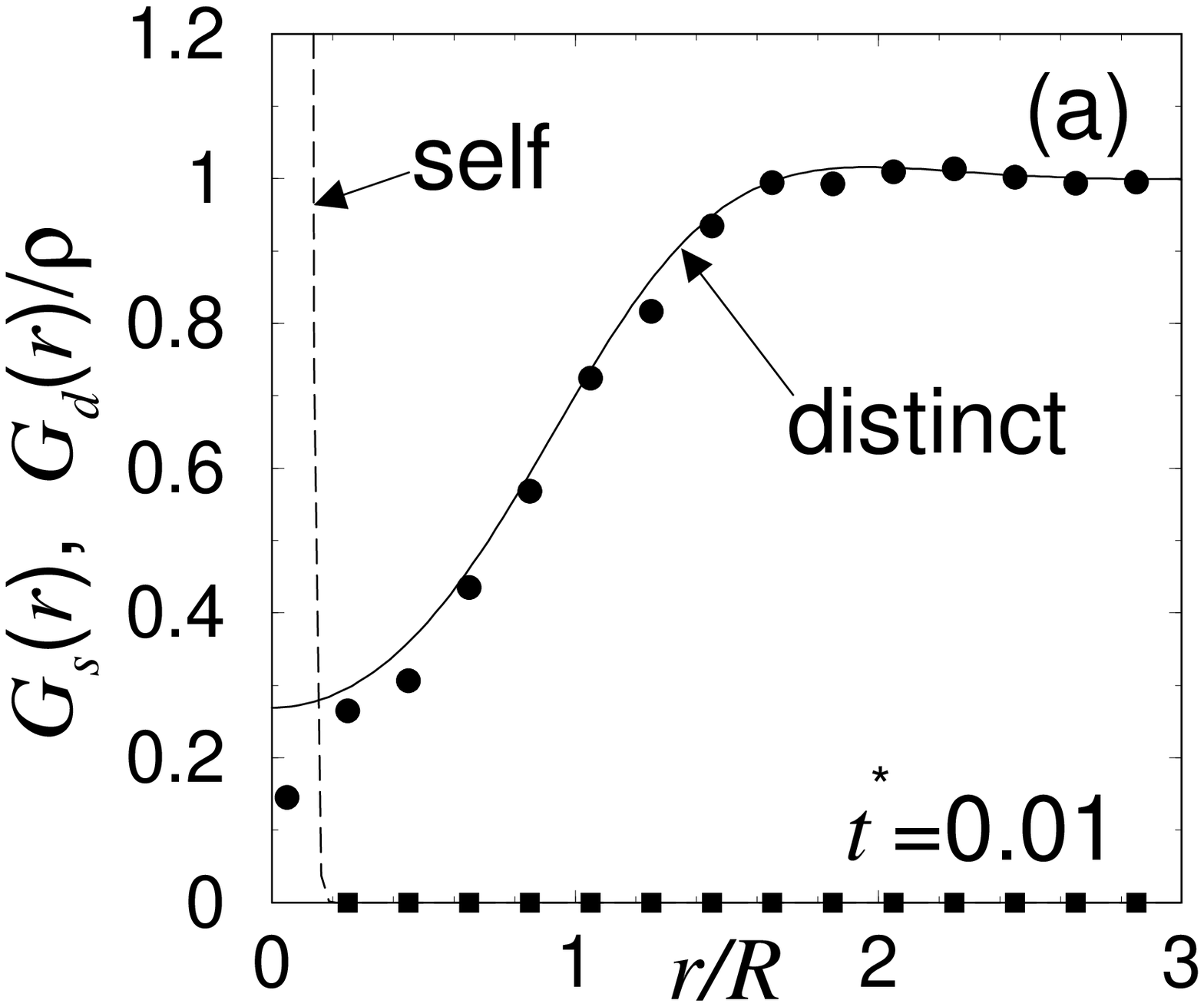}
\includegraphics[width=0.45\columnwidth]{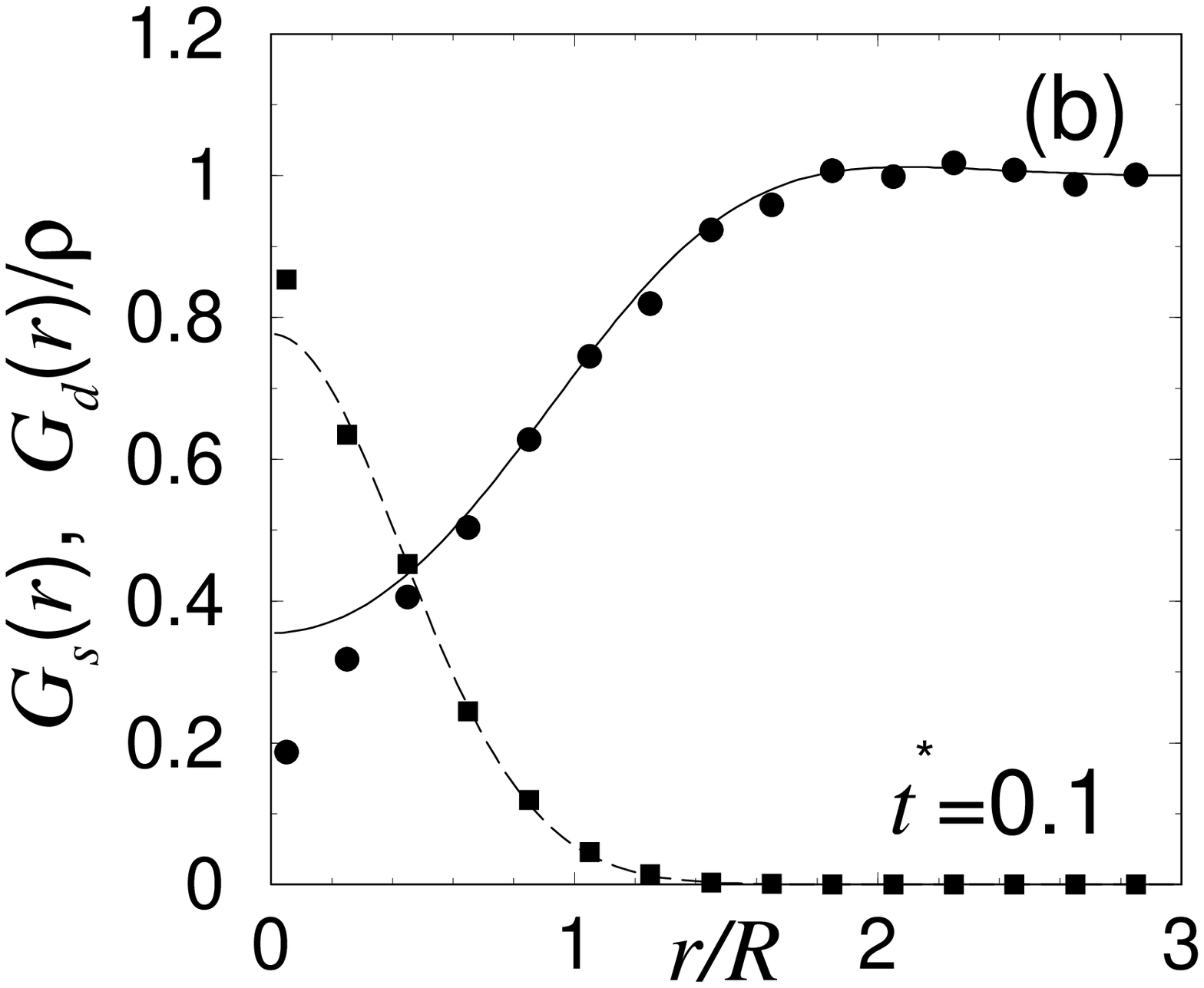}
\includegraphics[width=0.45\columnwidth]{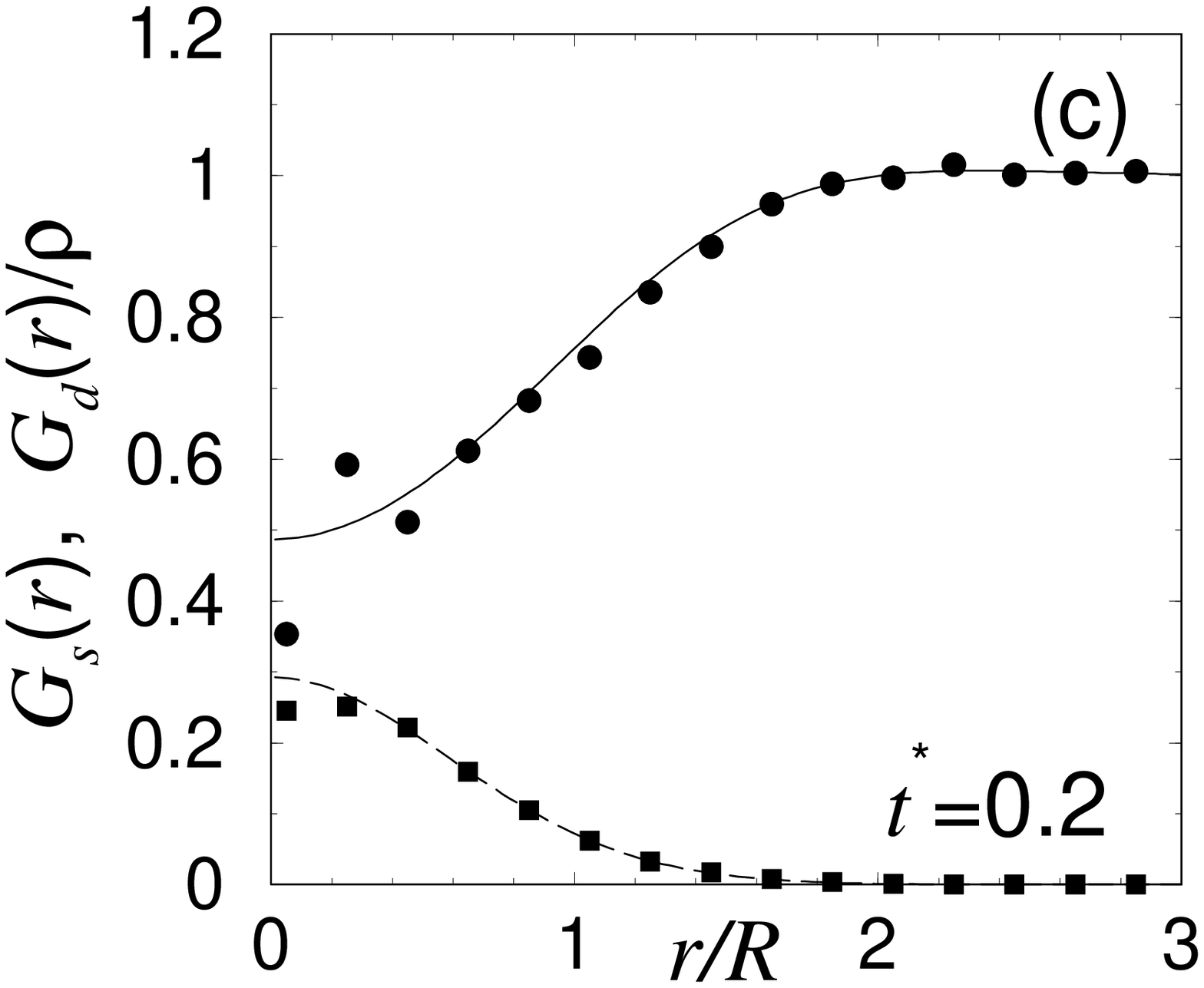}
\includegraphics[width=0.45\columnwidth]{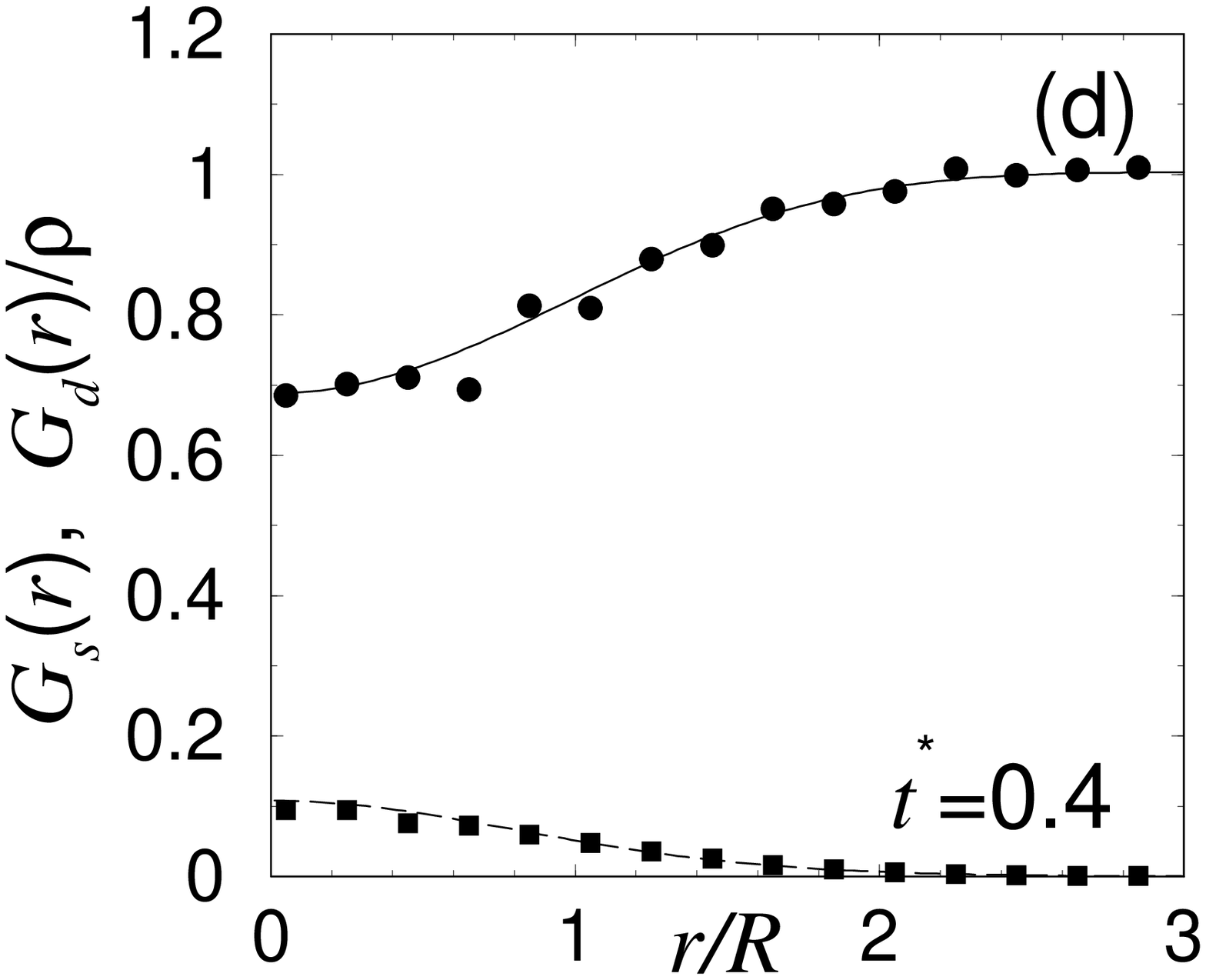}
\caption{\label{fig:van_hove} The self part $G_s(r,t)$ (dashed lines,
 squares) and scaled distinct part $G_d(r,t)/\rho$ (full lines,
 circles) of the van Hove correlation function for the GCM fluid with
 bulk density $\rho R^3=0.2263$ and $k_BT/\epsilon=0.5$, at times $t^*
 \equiv t/\tau_B=0.01,0.1,0.2$ and 0.4 (a,b,c,d, respectively), as a
 function of the (scaled) distance $r/R$. Shown are results from DDFT
 (lines) and BD simulations (symbols). }
\end{figure}

As a test case we consider the GCM, that describes the soft
interactions of polymer coils, star polymers, or dendrimers in
solution \cite{Likos}, and is defined by $V(r)=\epsilon
\exp(-r^2/R^2)$; $R$ is the radius of the particles and $\epsilon$ is
the energy cost for complete overlap of a pair of particles. We use a
simple mean field approximation, $F_{\rm ex}[\rho_s,\rho_d] =
\frac{1}{2}\sum_{i,j=s,d} \int d\rv d\rv' \rho_i(\rv)
V_{ij}(|\rv-\rv'|) \rho_j(\rv')$, which has been shown to be reliable
for this model \cite{litgaussians}, and integrate Eq.\
(\ref{eq:MT_DDFT}) numerically with the given initial conditions
forward in time. In Fig.~\ref{fig:van_hove} results are displayed for
both parts of the van Hove function at four different times for a
typical statepoint.  We also display results from Brownian dynamics
(BD) simulations of 300 particles, with a time step of $10^{-3}
\tau_B$, where $\tau_B=R^2/(\Gamma k_BT)$.  The very good agreement
between the DDFT and the BD simulation results attests to the validity
of our scheme.  Nevertheless, a remark about the more general
applicability of the DDFT in Eq.~(\ref{eq:MT_DDFT}) is in order. The
relevant dynamical variable is the root mean square particle
displacement $w(t)$, defined as the width of $G_s(r,t)$ [and hence of
$\rho_s(r,t)$] via
\begin{equation}
  [w(t)]^2=4 \pi\int_0^\infty {\mathrm d}r \, r^4 G_s(r,t).
  \label{eq:w}
\end{equation}
For $t \to \infty$, $w(t)^2 = 6 D_L t$, where $D_L$ is the long time
self diffusion coefficient \cite{Hansen06}.  However, for not too low
temperatures and $t \to \infty$, Eq.~(\ref{eq:MT_DDFT}) predicts that
$w(t)^2 = 6 D_S t$, where $D_S=k_BT /\Gamma$ is the {\em short time}
self-diffusion coefficient, and for the GCM $D_L \simeq D_S$
\cite{Archer17}. However, for particles with a hard core $D_L<D_S$ and
the DDFT in Eq.~(\ref{eq:MT_DDFT}) is not reliable for
$t\gg\tau_B$. To circumvent this problem one could use a DDFT such as
that of Ref.~\cite{Archer17}, which guarantees the correct long time
behaviour -- see also the approach of Ref.~\cite{schweizer06}.

Here we choose a different direction and investigate the free energy
landscape underlying Eq.\ (\ref{eq:MT_DDFT}) in more detail. We
consider the `fruit-fly' of liquid state physics, namely a
one-component system of hard spheres. This is described by the
Ramakrishnan-Yussouff (RY) approximation \cite{Hansen06,RY}:
$F_{\rm ex}[\rho_s,\rho_d] = Vf_{\rm ex}(\rho)
+ \sum_{i=s,d}f_{\rm ex}'(\rho)\int d\rv \Delta\rho_i(\rv)
- \frac{1}{2} \sum_{i,j=s,d} \int d\rv d\rv'
\Delta\rho_i(\rv)c_{ij}(|\rv-\rv'|)\Delta\rho_j(\rv')$, where
$f_{\rm ex}(\rho)$ is the bulk excess free energy per volume $V$,
$f_{\rm ex}'(\rho)=\partial f_{\rm ex}(\rho)/\partial \rho$,
$\Delta \rho_i(r)=\rho_i(r)-\rho_i$, with $\rho_d=\rho$, $\rho_s=0$,
and the Percus-Yevick approximation \cite{Hansen06} for the pair
direct correlation function $c_{ij}(r)$; we set $c_{ss}(r)=0$ to model
the absence of test particle self-interactions. Chosen here for its
simplicity, this approximation for $F_{\rm ex}[\rho_s,\rho_d]$ is
crude, but sufficient to explore qualitative behaviour.  We use
equilibrium DFT to elucidate a pathway that closely resembles that of
the full DDFT by minimising the free energy functional \cite{Evans92},
i.e.\ solve $\delta F/\delta \rho_i(\rv) = \mu_i$ for $i=s,d$ where
$\mu_i$ are Lagrange multipliers to fulfill $\int \dr \rho_s(\rv)=1$
and $\int \dr(\rho_d(\rv)-\rho) = \rho \int \dr (g(r)-1) = {\rm
const}$.  As a constraint we control $w$ via an associated Lagrange
multiplier, $\lambda$, which is formally equivalent to treating an
auxiliary external field acting on component $s$ with the simple
harmonic form $V_{\rm ext}^{(s)}(r)=\lambda r^2/\sigma^2$.

\begin{figure}
\includegraphics[width=1.\columnwidth]{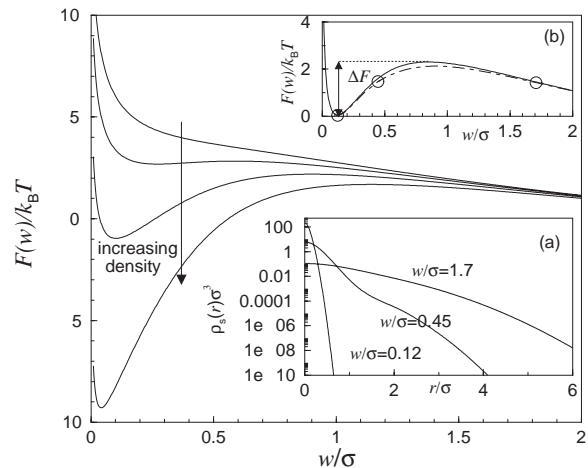}
\caption{\label{fig:F} The scaled free energy $F/k_BT$ as a function
  of the root mean square particle displacement $w$ [see
  Eq.~(\ref{eq:w})], calculated with the RY functional for hard
  spheres at densities $\rho\sigma^3=0.6,0.7,0.8$ and 0.9 (as
  indicated by the arrow) and shifted such that $F(w=3)=0$.  (a) Self
  part of the van Hove function on a logarithmic scale as a function
  of $r/\sigma$ for $\rho\sigma^3=0.78$ and three typical widths
  $w$. Note the pronounced non-Gaussian character for $w/\sigma=0.45$.
  (b) Comparison of results for $F/k_BT$ as a function of $w/\sigma$
  from free minimisation (dot-dashed line) and a Gaussian
  parametrisation for $G_s$ (solid line) for $\rho\sigma^3=0.78$. Also
  shown is the free energy barrier $\Delta F$ between minimum and
  maximum (arrow) and the values of $w/\sigma$ (circles) for which
  results are shown in (a).}
\end{figure}

By calculating $F$ as a function of $\lambda$ (and hence $w$) for a
given statepoint, we may determine the free energy landscape that
governs the time evolution of $G(r,t)$. We find that at low densities
$F(w)$ decreases monotonically with $w$, but as $\rho$ is increased
above a threshold, $\rho>\rho^*$, $F(w)$ becomes non-monotonic and
develops a barrier; the barrier height $\Delta F$, i.e.\ the
difference between the minimum and maximum, grows upon increasing
$\rho$, see Fig.~\ref{fig:F}.  While for both small and large values
of $w$, $\rho_s(r;\lambda)$ is nearly Gaussian, for intermediate
values, particularly when $\rho \gtrsim \rho^*$, $\rho_s(r;\lambda)$
has pronounced non-Gaussian characteristics; Fig.\ \ref{fig:F}a shows
results for $\rho_s(r)$ for three typical values of $w$. For states
that correspond to barrier crossing $\rho_s(r)$ exhibits a pronounced
shoulder, taken as a signature of dynamical heterogeneity
\cite{kegel00,kob97}.  However, assuming the Gaussian form for all
values of $w$, as in the main plot of Fig.~\ref{fig:F}, does not
affect $F(w)$ strongly, and only slightly overestimates $\Delta F$
(see Fig.~\ref{fig:F}b for a comparison of results from free
minimisation and from the Gaussian parametrisation).  The emergence of
a barrier in $F(w)$ leads to a trapping in the DDFT for $\rho>\rho^*$
whereby, as $t \to \infty$, $w(t)\to w_0 < \infty$, suggesting that
the system is non-ergodic. However, in reality, as long as the barrier
is sufficiently small, it can eventually be overcome.  Our expression
for $F_{\rm ex}$ is approximate and neglects some contributions to the
free energy due to fluctuations \cite{Evans92,reguera04}, so the
particular DDFT in Eq.~(\ref{eq:MT_DDFT}) does not describe the
barrier crossing; one could include a stochastic noise term, following
Ref.\ \cite{schweizer06}, to take this into account.

We may use our results to estimate when the hard sphere fluid becomes
trapped in a glassy state by considering the time $\tau$ it takes to
cross the barrier, which scales as $\tau\sim e^{\Delta F/k_BT}$
\cite{schweizer06,schweizer03and05}. We estimate the location of the
glass transition by determining the state point at which $\Delta F$
reaches a particular value; as a rough criterion we take
$10k_BT$. From the results in Fig.~\ref{fig:F} one would expect this
to occur around $\rho_g \sigma^3 \simeq 0.9$, which is somewhat below
the result $\rho_g\sigma^3 \simeq 1.1$ found experimentally for
colloidal hard spheres \cite{vanMegenUnderwoodPRL1993}. We attribute
this under-estimation to our choice of simple (RY) free energy
functional; use of a more reliable functional \cite{FMT} may improve
results.

We emphasise that our approach is distinct from theories describing
the glass transition as freezing into an aperiodic lattice
\cite{Wolynes}; we do not find a thermodynamic phase transition. There
is some similarity between our philosophy and that underlying the
successful `non-classical theory of nucleation', where DFT is used to
calculate the free energy barrier to nucleating a liquid droplet in
the oversaturated gas \cite{OxtobyEvansJCP1988}. The results presented
in Fig.~\ref{fig:F} bear much similarity to those presented in recent
work by Schweizer and Saltzman \cite{schweizer06,schweizer03and05} who
also construct a theory for the free energy of a single particle in
the cage of its neighbours.  One key difference is that we calculate
the free energy of the entire system, including contributions from
$\rho_d(r)$ and $\rho_s(r)$, whereas their theory is for the free
energy of the test particle only. Furthermore our approach is readily
applicable to inhomogeneous systems.

\begin{figure}
\includegraphics[width=0.95\columnwidth]{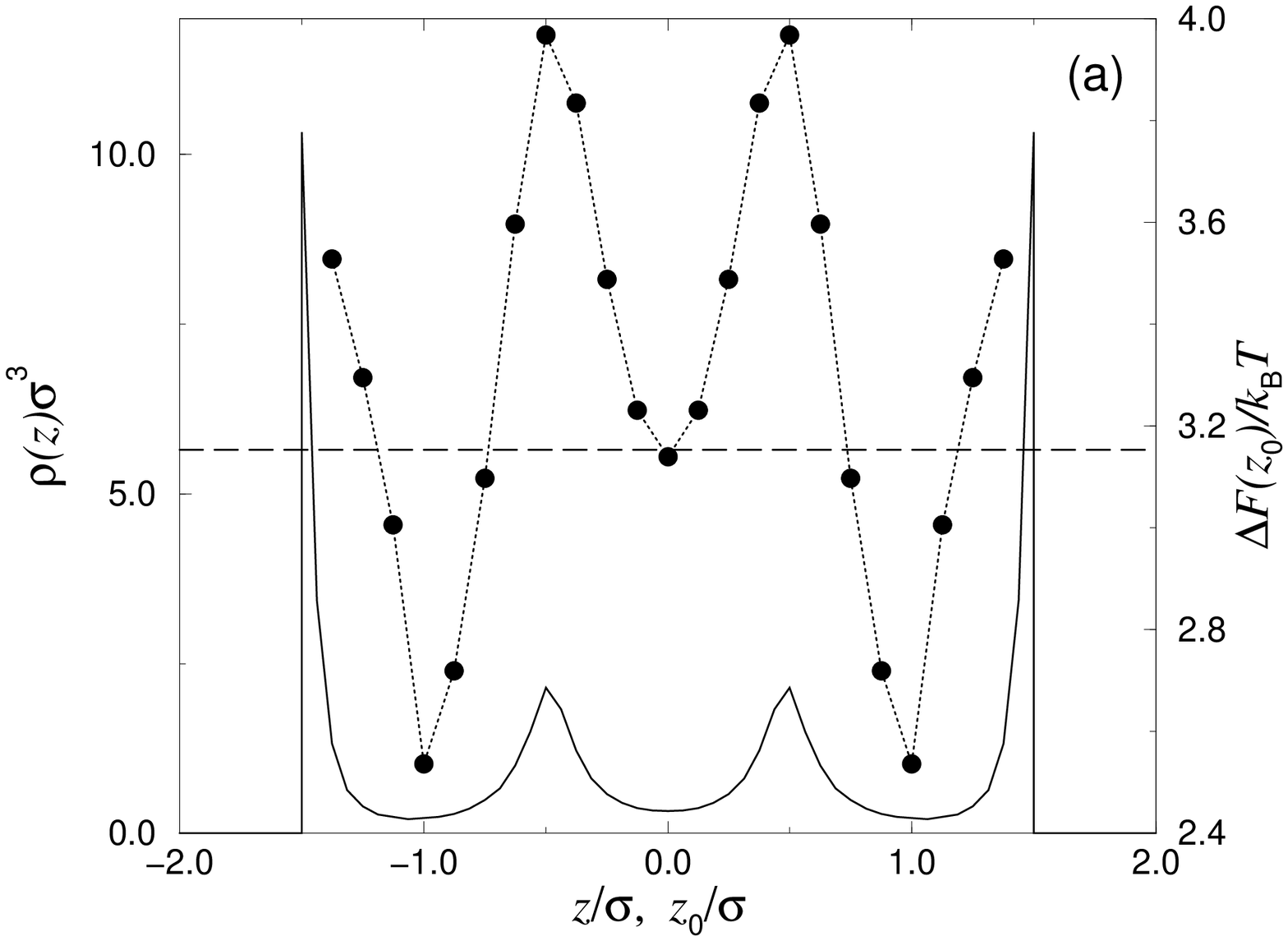}
\includegraphics[width=1.\columnwidth]{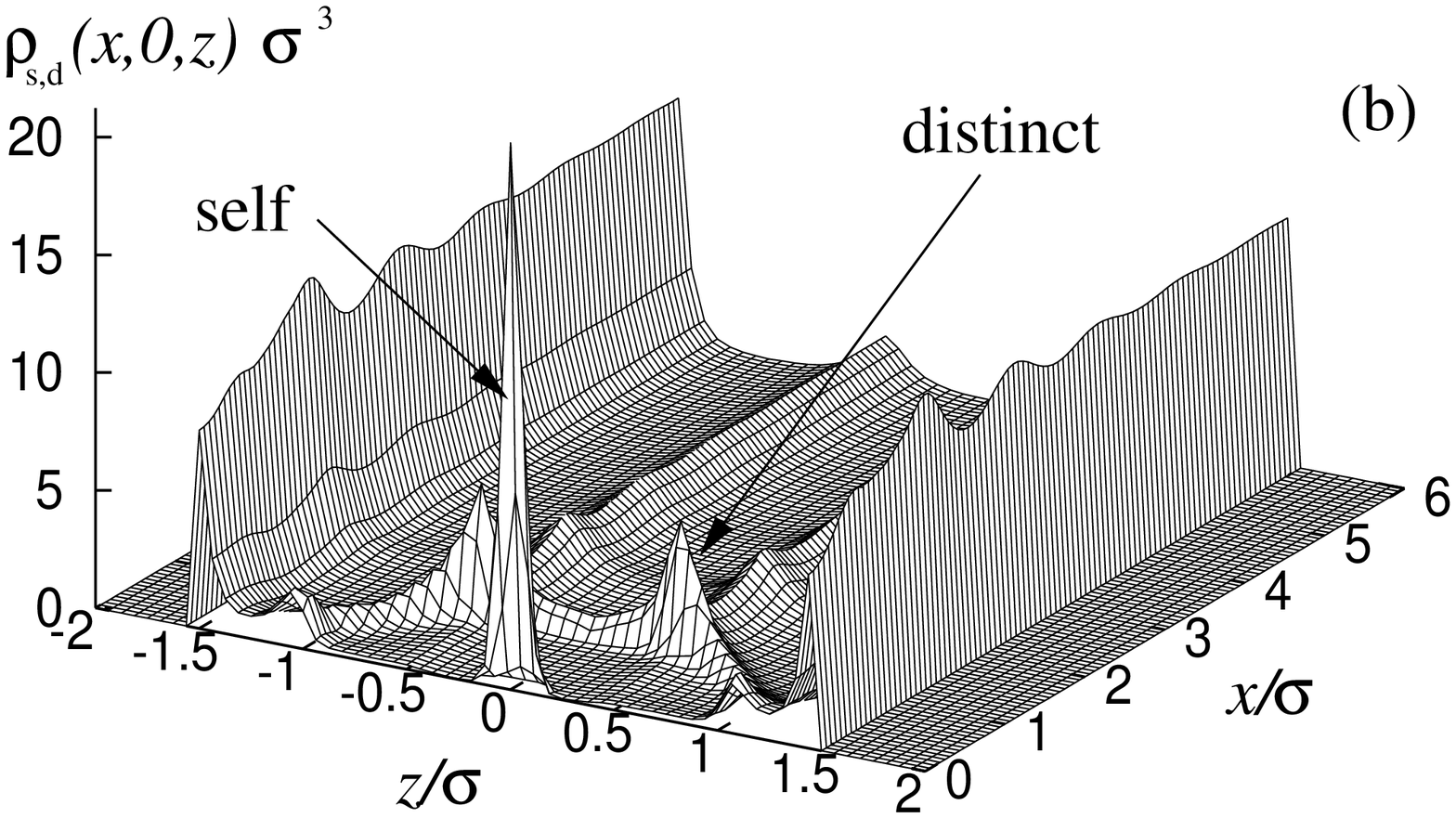}\vspace{-5mm}
\caption{\label{fig:confined} Correlation functions for hard spheres
 confined between parallel plates of separation distance $H=4\sigma$.
 (a) Density profile $\rho(z)$ (full line) as a function of $z/\sigma$
 and free energy barrier $\Delta F(z_0)$ (symbols; the dotted line is
 to guide the eye). Also shown is the corresponding bulk value of
 $\Delta F$ (dashed line).  (b) The (scaled) self part
 $\rho_s(\rv)\sigma^3/25$ and distinct part $\rho_d(\rv)$ of the van
 Hove function as a function of the scaled coordinates perpendicular
 and parallel to the wall, $z/\sigma$ and $x/\sigma$, respectively,
 evaluated at the minimum of the free energy landscape for bulk
 density $\rho\sigma^3=0.8$.  }
\end{figure}

We have investigated how the dynamics of hard spheres is affected by
confinement between two planar parallel hard walls separated by a
distance $H$, described by an external potential $V_{\rm ext}(z)$ that
vanishes for $-H/2<z<H/2$, and is infinite otherwise; $z$ is the
coordinate perpendicular to the walls.  The presence of the walls
causes the one-body density, $\rho(z)$, to be oscillatory with maximal
value at contact with the walls; see Fig.\ \ref{fig:confined}a for
results for a system with $H=4\sigma$ in chemical equilibrium with a
bulk of density $\rho\sigma^3=0.8$. Due to symmetry the van Hove
function depends on the initial position $z_0$ of the test particle,
but not on its initial lateral coordinates, taken as $x_0=y_0=0$. We
have used equilibrium DFT to determine $\rho_d(\rv)$, assuming
$\rho_s(\rv)$ to be a Gaussian (recall that in bulk this assumption
made little difference to the value of $F$).  As an example we show in
Fig.\ \ref{fig:confined}b results for $\rho_s(\rv)$ and $\rho_d(\rv)$
for the case where the test particle was initially in the midplane,
$z_0=0$. We choose conditions (see below) where $\rho_s(\rv)$ is still
peaked around $\rv=0$; $\rho_d$ exhibits structure due to packing
effects caused by the walls, as well as by packing around the test
particle. Quite striking is the appearance of the hexagonal shape of
the first coordination shell; for large planar distances
$\rho_d(z,x\to\infty,y=0)=\rho(z)$.  Similar to the bulk case we find
a free energy barrier $\Delta F$ that now depends on $z_0$ (the
profiles in Fig.\ \ref{fig:confined}b are shown at the minimum of
$F$.)  Fig.\ \ref{fig:confined}a shows strong oscillations of $\Delta
F(z_0)$ around its value in bulk; the oscillations are in phase with
the oscillations of $\rho(z)$.  Since the barrier hopping time $\tau
\sim e^{\Delta F/k_BT}$, the particles in the layers (i.e.\ where
$\rho(z_0)$ is large) are less mobile than those between the
layers. For the case in Fig.\ \ref{fig:confined}b the particles at $z=
\pm 0.5 \sigma$ have a mobility ($\sim 1/\tau$) that about 50\% of the
bulk value. This prediction can be tested by experiment with confined
colloidal hard spheres or with computer simulations.

In conclusion we have presented a general framework for calculating
the van Hove two body correlation function both in bulk and in
inhomogeneous fluids. We have proposed a method for determining the
free energy landscape for a particle in the cage of its neighbours and
have demonstrated the applicability to inhomogeneous systems, where
e.g.\ mode coupling theory \cite{Hansen06} cannot be easily applied
\cite{Archer17}.

We thank R.\ Evans, M.\ Fuchs, J.M.\ Brader and R.\ van Roij for
valuable discussions. A.J.A. and P.H. are grateful for the support of
EPSRC and M.S.\ thanks the DFG for support through the SFB TR6/D3.

\end{document}